# Photoinduced ultrafast dynamics of local nematicity and lattice distortions in FeSe crystals

T. Konstantinova[1,2], L. Wu[1], M. Abeykoon[1], R. J. Koch[1], A. F. Wang[1, §], R. K. Li[3], X. Shen[3], J. Li[1], J. Tao[1], I. A. Zaliznyak[1], C. Petrovic[1], S. J. L. Billinge[1,4], X. J. Wang[3], E. S. Bozin[1] and Y. Zhu[1,2]

[1] Brookhaven National Laboratory, Upton, NY, USA 11973
[2] Stony Brook University, Stony Brook, NY, USA 11794
[3] SLAC National Accelerator Laboratory, Menlo Park, CA, USA 94025
[4] Columbia University, New York, NY 10027, USA
[§] Present address: School of Physics, Chongqing University, Chongqing 400044, China

Formation of electronic nematicity is a common thread of unconventional superconductors. In iron-based materials, the long-range nematic order is revealed by the orthorhombic lattice distortion, which importance is a highly controversial topic due the small magnitude of the distortion. Here, we study the local crystal structure of FeSe and its interaction with electronic degrees of freedom using ultrafast electron diffraction, x-ray pair distribution function analysis, and transmission electron microscopy and find a significant lattice response to local nematicity. The study demonstrates how local lattice distortions, which exist even at temperatures above the nematic phase transition, can be released by photoexcitation, leading to an enhancement of the crystalline order. The observed local atomic structures and their out-of-equilibrium behavior unravel a sophisticated coupling between the lattice and nematic order parameter in FeSe.

FeSe is the simplest iron-chalcogenide superconducting compound. In a common trend with other Fe-based superconductors (FBSC), at low temperature FeSe undergoes a transition to the nematic electronic state, deemed a precursor of superconductivity, which is accompanied by a weak change in the average crystal symmetry from tetragonal P4/nmm to orthorhombic Cmma group. The origin of the nematicity is still under debate[1], with spin[2], orbit[3] and charge fluctuations[4] considered as main mediators. Antiferromagnetic order, which usually closely follows the transition to the nematic state in iron pnictides supports the spin scenario as a leading contender for the nematic order. FeSe, however, lacks a long range magnetic order whereas only local spin correlations are observed[5]. Hence, FeSe presents unique opportunity for the investigation of the formation of nematicity and its relation to non-conventional superconductivity.

The connection between the crystal lattice and electronic nematicity in FBSC is often neglected because of apparently insignificant change of the unit cell parameters in the nematic phase. In FeSe, a tetragonal-to-orthorhombic transition leads to only 0.5% distortion in the *ab*-plane[6]. Nevertheless, there is a growing evidence of coupling between the lattice and electronic degrees of freedom in this material. It includes sensitivity of superconducting temperature $T_C$ and structural transition temperature $T_S$ to pressure[7] and lattice strain[8], isotope effect[9], optical phonon anomaly[10], phonon softening[11] and enhancement of $T_C$ in a single layer FeSe on $SrTiO_3$ substrate[12,13] and in FeSe crystals with spacer layers[14].

In this work, we present detailed study of the local crystal structure of FeSe using pair distribution function analysis of x-ray powder diffraction (XPD) and transmission electron microscopy (TEM) and investigate the dynamics of structural changes following the photoinduced melting of the nematic order using ultrafast electron diffraction (UED). Our experiments reveal surprising increase of crystallinity upon melting of low-symmetry local lattice distortions, which we associate with local nematicity. These distortions are present in both long-range nematic (orthorhombic) and normal (tetragonal) states; however, their correlation increases below $T_S$. Upon photoexcitation, the distortions are released at a rate that is determined by the presence of the residual long-range nematic order, implying direct coupling between nematicity and the local crystal structure. The observations can explain the electronic nematic fluctuations detected[15-18] above $T_S$.

Single FeSe crystals were grown by the chemical vapor transport method using a eutectic mix of the KCl and $AlCl_3$ as the transport agent[19,20]. UED measurements were performed at MeV-UED setup at SLAC Accelerator National Laboratory. The details of the UED setup are described elsewhere[21]. XPD measurements for PDF analysis were performed at XPD-beamline of National Synchrotron Light Source–II at Brookhaven National Laboratory. High-resolution XPD data were obtained at the 11-BM beamline at the Advanced Photon Source at Argonne National Laboratory. TEM measurements including diffraction and imaging were performed at BNL using 200 keV JEOL ARM 200 CF Microscope with a probe and an imaging aberration corrector.

UED provides information about the lattice dynamics in the system driven out-of-equilibrium with a pump laser pulse probed by an electron beam. We use 1.55eV-60fs photon pulse to excite electronic transitions in FeSe samples at temperatures from 27 K to 300 K. To get information about the lattice response we focus on the intensity changes of the Bragg reflections that can be related the lattice symmetries with different length of scattering vectors. Typical intensity dynamics of *<200>*, *<020>* and *<400>*, *<040>* at 27 K are shown in Fig. 1 (we use Cmma symmetry for indexing peaks, unless stated otherwise). During the first 5 picoseconds (ps) the intensity of all observable peaks goes down by a few percent of the initial equilibrium values [Fig. 1(a, b)]. At this timescale, the intensity dynamics can be fitted with a single exponential decay with a time constant of 1.5-2 ps. Such behavior is consistent with the energy transfer from the excited electrons to the lattice through electron-phonon coupling, leading to increase of atomic Debye-Waller factors. Similar time constants were observed for the initial recovery of electronic states in the time-resolved reflectivity experiments [18] for the same material and were also attributed to the electron-phonon coupling.

Beyond the first 5 ps the dynamics of the Bragg reflections are rather unusual. The intensities of the *<hk0>* Bragg peaks with $h+k=4n+2$ continue to drop for about 50 ps and then slowly recover. On the other hand, the intensities of the rest of the peaks ($h+k=4n$) increase well above the initial values within the same 50 ps, before recovery. The rate of the lattice changes in the 5 ps to 50 ps interval is similar to the rate of the photoinduced orthogonal-to-tetragonal phase transition in $BaFe_2As_2$ observed[22] with time-resolved X-ray diffraction. However, structural changes associated with the same transition cannot lead to the observed intensity variations in the present experiment.

Consider the tetragonal *220* peak that splits into the orthorhombic *400* and *040* peaks. Such splitting is too small to be observed by UED since the experimental widths of the Bragg peaks are several times larger than the expected splitting. Assuming the high symmetry positions of Fe and Se atoms obtained from the Rietveld refinements, the structure factors of these peaks, which determine their intensities in a thin sample, have identical form,

$$SF_{220}^{tetr} = SF_{040}^{orth} = 4f_{Se}\exp(-B_{Se}) + 4f_{Fe}\exp(-B_{Fe}) \qquad (1)$$

where $f_{Se}(f_{Fe})$ and $B_{Se}(B_{Fe})$ are the atomic form factor and the Debye-Waller factor for Se (Fe) atoms respectively. From Eq. (1) one can see that not only the transition between the two phases does not lead to an intensity change, but that no modification of atomic positions could increase the intensity of *<040>* (and other $h+k=4n$) peaks because for those reflections the electrons already scatter in phase from all atoms in the unit cell. An apparent intensity increase suggests that some lattice distortions *preexist* at equilibrium, yielding a reduced Bragg intensity compared to the ideal structure factor of Eq. (1). In such case, a photo-driven release of these distortions results in the intensity rise.

A closer look at the shape of the diffraction peaks provides additional information about the lattice dynamics. The inset in Fig. 1(c) shows the intensity profile integrated along the [110] direction. Changes in the profile [Fig. 1(d)] at time delays around +55 ps and beyond +300 ps show that the integrated intensity of $h+k=4n$ peaks increases. However, changes at peaks' centers are different from changes at peaks' tails. A separation of an individual peak's profile into a narrow Gaussian part, corresponding to the long-range crystal order, and a wider Lorentzian part, corresponding to a short-range order [SI] shows that the lattice dynamics involve three major steps. Firstly, the photoinduced atomic vibrations lead to the decrease of the Bragg peaks' intensity, which is transferred to the thermal diffuse background. Secondly, the release of the pre-existing distortions, which in the absence of the photoexcitation give rise to a broad diffuse scattering near q=0, induces the re-crystallization of the high-symmetry phase, i.e. causes changes in the average crystal structure (long-range order) by moving atoms to more symmetric positions. This displacive process leads to increase(decrease) of structure factors for $h+k=4n$ ($h+k=4n+2$) peaks. The behavior of the Gaussian component is determined by the combination of vibrational and displacive effects. Thirdly, melting of the local distortions also creates tiny domains of high-symmetry phase, increasing the intensity of the Lorentzian component. The size of the domains determines the width of the Lorentzian component and can be estimated around 15-20 Å.

The crystal lattice in S-doped samples photoexcited under comparable conditions demonstrates similar response [SI], pointing out that the distortions are common for at least a part of the FeSe$_{1-x}$S$_x$ phase diagram. To understand the nature of these pre-existing local distortions, breaking the lattice symmetry at equilibrium, we turn to static techniques such as XPD and TEM.

Atomic displacements, corresponding to the bond disparity of 0.1 Å have been observed in Fe$_{1+y}$Te, where they were attributed to a long-range ferro-orbital ordering[23]. To search for similar atomic displacements in FeSe, we performed Pair Distribution Function (PDF) analysis of the XPD data. PDF contains information about both long-range order and local imperfections, which is inferred from the powder diffraction pattern. Fig. 2(a) shows PDF data at T = 84 K together with the fit to an orthorhombic model, obtained from the Rietveld refinement. Whereas the model describes the data well at large interatomic distances $r$, for $r < 10$ Å there is a notable misfit. The misfit indicates that there is a disparity between the local and average atomic structure and corroborates the assumption that lattice distortions are present at equilibrium.

Remarkably, a pronounced misfit to the tetragonal model at small inter-atomic distances is also present at 300 K [Fig. 2(c)]. Thus, the local lattice distortions also exist in the tetragonal phase. The deviation from the tetragonal model is most pronounced for the lattice repeating peak at $r$=3.8 Å and rapidly fades at larger $r$, indicating short correlation length. The information about the distortions in PDF comes not from Bragg peaks, but from diffuse scattering. It agrees with the UED observations, where melting of the distortions involves intensity transfer from the diffuse background centered at $q$=0 to locations at or near Bragg peaks. The exact structure of the distortions is the subject of a separate publication.

Whereas XPD provides structural information averaged over multiple lattice domains, TEM is a local probe and presents an opportunity to look at individual domains and to reconstruct the details that could be missed upon averaging under a large probe. The results of our TEM measurements of FeSe samples are shown in Fig. 3. In agreement with the previous studies on FeSe[6] and LaOFeAs[24], *<110>* peaks forbidden by Cmma symmetry appear in the diffraction pattern below $T_S$ (at T=88 K) [Fig. 3(a,b)] whereas they are not seen at the same sample area at T=300 K. The peaks indicate that the crystal symmetry below $T_S$ is lower than Cmma. Such peaks were not detected in the XPD measurements [SI].

Fig. 3(c) shows a High Resolution TEM (HRTEM) image obtained at 300 K with a smaller field view than is used for the diffraction measurements. Fourier analysis of such images [Fig. 3(d)] reveals nonuniformly distributed regions whose diffractograms have a pair of forbidden peaks (either $\bar{1}10$ and $1\bar{1}0$ or $110$ and $\bar{1}\bar{1}0$), or a full set of four *<110>* peaks in addition to the peaks allowed by Cmma or P4/nmm symmetry [SI]. Yet other regions have only allowed peaks. Appearance of the peaks in either of the two diagonal directions in diffractograms can be explained by presence of domains with $C_2$ symmetry in the *ab*-plane, which are rotated by 90° with respect to each other. The difference in the forbidden peaks' intensities along the perpendicular directions of the large area diffraction at 88 K [SI] also supports the idea of rotated domains with $C_2$ symmetry, which are nonequally present in the probed volume below $T_S$. Additionally, a Fourier transform of a scanning TEM image also shows the forbidden peaks [SI]. These observations corroborate that even at 300 K the sample has regions with the broken tetragonal symmetry, where either or both atoms in the unit cell are displaced from the high symmetry positions, leading to the atomic bond disparity. The disparity agrees with the misfit of the PDF model described above. Neutron powder diffraction experiments in other FBSC have also observed[25,26] local structures that are different from the average ones.

The photoinduced FeSe lattice dynamics at different temperatures provide important information about changes in the system across the nematic phase transition. As shown in Fig. 4(a), the relatively fast (within 50 ps) increase of *<080>*, *<800>* peaks intensity, corresponding to release of the distortions, is only observed at temperatures below $T_S$. Above $T_S$ the intensity rises as well, however, at a much slower pace (within 400 ps). The photoinduced increase of intensity above $T_S$ agrees with the presence of local nematic distortions observed with x-ray and TEM. Remarkably, the relative intensity at 1 nanosecond delay seems to be independent of temperature. For dynamics of other peaks refer to the Supplemental Information.

Fluence dependence [Fig. 4(b)] of the lattice dynamics at 27 K also reveals switching between the fast and the slow regimes. The fast component is observed only at fluences below 2.2 mJ/cm$^2$. The value of the maximum intensity firstly growth with fluence and then drops above 1.9 mJ/cm$^2$. Above 2.2 mJ/cm$^2$ the lattice response is the same as at high sample temperature, i.e. only the slow component is observed. This can be explained by sample heating. Based on the sample characteristics[27,28], if all the absorbed energy is converted to heat the threshold fluence corresponds to a temperature increase of 75 K, which is close to 64 K difference between the sample temperature and $T_S$. Thus, the process leading to the fast increase of the peaks' intensity proceeds only in the presence of a (partial) long-range nematic order. When the order is destroyed, either by temperature or through above-the-threshold photoexcitation, the slow process governs the lattice dynamics. The slow process is also present at low temperature – low fluence regime. Figure 4(c) shows that the point of the maximum intensity shifts to the longer time with increased fluence, reflecting the increased impact of the slow process. Thus, the slow and the fast responses "compete" with each other: as the laser fluence (or sample temperature) increases the slower process becomes more pronounced and finally dominant.

It is often believed that weak orthorhombicity of the unit cell is the only result of coupling between the electronic nematic order and the lattice in FeSe superconducting family. Our observations reveal an additional connection, established via atomic bonds' distortions that lower the local lattice symmetry. Such distortions are present in orthorhombic nanodomains at temperatures both below and above $T_S$ and correspond to local nematic fluctuations, consistent with previous observations[15,16,18]. Their correlation length increases as domain size grows on cooling, leading to the percolative three-dimensional ordering below $T_S$. This transition and the presence of the uncorrelated low-symmetry domains both below and above the ordering temperature agrees with the theoretical predictions of the anisotropic random field Ising model (ARFIM), which was argued to describe phase transitions with the discrete two-fold symmetry breaking in layered systems[29,30].

A comparison of the photoinduced nonequilibrium lattice dynamics in FeSe and the previously observed dynamics of the electronic degrees of freedom provides important insights into the role of the distortions in formation of the nematic phase. It was shown[18,31] that the long-range electronic nematic order melts within few hundred femtoseconds upon photoexcitation. Yet no changes in the lattice, except for the mild onset of the heating due to electron-phonon coupling, are observed at this time scale. Release of the distortions, i.e. motions of atoms towards their high symmetry positions, proceeds with a much slower pace, meaning that the amplitude of the distortions is more robust against photoexcitation than the long-range nematic order. Thus, the atomic off-symmetry displacements at low temperature are not a mere consequence of the nematic phase. Whereas the amplitude of the distortions is unchanged, the TEM and XPD data imply that the coherence across the distorted regions gets lost during melting of the nematic order.

A notable feature of the nonequilibrium lattice dynamics is the threshold excitation fluence at low temperatures, corresponding to the energy needed to completely melt the long-range nematic order. Below the threshold fluence, i.e. when a partial nematic order parameter is still present after the excitation, the rate of the distortion release is relatively fast and matches the rate of the recovery of the electronic nematicity[18]. Excitation above the threshold fluence leads to a complete melting of the nematic order parameter by 'overheating' the sample and results in a slow relaxation of the lattice distortions, same as observed at temperatures above $T_S$. The presence of the long-range nematic order as a prerequisite for the fast distortion release implies coherent atomic rearrangements leading to growth of domains with high lattice symmetry, which, however, are unstable and partially split back into smaller domains within next few hundred picoseconds. Change of the distortion amplitude in the slow process depends only on the excitation fluence and not on the sample temperature. Presumably, the slow process reflects the gradual equation of distinct atomic bonds, which transforms some of the pre-existing low-symmetry nematic nanodomains into high-symmetry parent phase.

In summary, our integrated study reveals nanodomains with local low-symmetry lattice distortions in FeSe that couple to the electronic degrees of freedom, thus exposing the short-range fluctuations of the nematic state. Using UED, we observe ultrafast melting of these nematic nanodomains following a femtosecond laser pulse and the concomitant ultrafast lattice ordering of the high-symmetry parent phase leading to a surprising increase of coherent Bragg scattering. The pre-existing local distortions are present at equilibrium both in the absence of long-range nematic order at 300 K and in the ordered phase and their ultrafast dynamics can be understood from the anisotropic random field Ising model theory[29,30]. ARFIM phase diagram predicts the domains of both low- and high-symmetry phases below and above the percolative phase transition[32,33]. Redistribution of the relative population of the two phases occurs via fast motion of the domain boundaries, thus naturally explaining the ultrafast structural response. Our study sheds new light into nematic order in the system and stimulates further theoretical development towards full explanation of the nematicity in FeSe.

Authors thank A. F. Kemper, L. Classen, A. V. Chubukov, R. M. Konik and A.M. Lindenberg for useful discussions and Saul Lapidus for XPD measurements at APS. This research used 28-ID-1 (PDF) beamline of the National Synchrotron Light Source II and JEOL ARM 200 CF Microscope at BNL. Work at Brookhaven National Laboratory was supported by U.S. Department of Energy (DOE), Office of Science, Office of Basic Energy Sciences (BES), Materials Science and Engineering Division under contract DE-SC0012704. The UED work was performed at SLAC MeV-UED, which is supported in part by the DOE BES Scientific User Facilities Division Accelerator & Detector R&D program, the LCLS Facility, and SLAC under contract Nos. DE-AC02-05-CH11231 and DE-AC02-76SF00515.

[1]      R. M. Fernandes, A. V. Chubukov, and J. Schmalian, Nat Phys **10**, 97 (2014).

[2]      C. Fang, H. Yao, W. F. Tsai, J. P. Hu, and S. A. Kivelson, Phys Rev B **77** (2008).

[3]      S. H. Baek, D. V. Efremov, J. M. Ok, J. S. Kim, J. van den Brink, and B. Buchner, Nat Mater **14**, 210 (2015).

[4]      P. Massat *et al.*, P Natl Acad Sci USA **113**, 9177 (2016).

[5]      T. Imai, K. Ahilan, F. L. Ning, T. M. McQueen, and R. J. Cava, Phys. Rev. Lett. **102**, 177005 (2009).

[6]      T. M. McQueen, A. J. Williams, P. W. Stephens, J. Tao, Y. Zhu, V. Ksenofontov, F. Casper, C. Felser, and R. J. Cava, Phys Rev Lett **103** (2009).

[7]      S. Medvedev *et al.*, Nat Mater **8**, 630 (2009).

[8]      S. H. Baek, D. V. Efremov, J. M. Ok, J. S. Kim, J. van den Brink, and B. Buchner, Phys Rev B **93** (2016).

[9]      R. Khasanov, M. Bendele, K. Conder, H. Keller, E. Pomjakushina, and V. Pomjakushin, New J Phys **12** (2010).

[10]     M. Nakajima, K. Yanase, F. Nabeshima, Y. Imai, A. Maeda, and S. Tajima, Phys Rev B **95** (2017).

[11]     K. Zakeri, T. Engelhardt, T. Wolf, and M. Le Tacon, Phys Rev B **96** (2017).

[12]     J. F. Ge, Z. L. Liu, C. H. Liu, C. L. Gao, D. Qian, Q. K. Xue, Y. Liu, and J. F. Jia, Nat Mater **14**, 285 (2015).

[13]     W. W. Zhao, M. D. Li, C. Z. Chang, J. Jiang, L. J. Wu, C. X. Liu, J. S. Moodera, Y. M. Zhu, and M. H. W. Chan, Sci Adv **4** (2018).

[14]     M. Burrard-Lucas *et al.*, Nat Mater **12**, 15 (2013).

[15]     P. S. Wang, P. Zhou, S. S. Sun, Y. Cui, T. R. Li, H. C. Lei, Z. Q. Wang, and W. Q. Yu, Phys Rev B **96** (2017).

[16]     Y. C. Wen, K. J. Wang, H. H. Chang, J. Y. Luo, C. C. Shen, H. L. Liu, C. K. Sun, M. J. Wang, and M. K. Wu, Phys Rev Lett **109** (2012).

[17]     K. Nakayama, Y. Miyata, G. N. Phan, T. Sato, Y. Tanabe, T. Urata, K. Tanigaki, and T. Takahashi, Phys Rev Lett **113** (2014).

[18]     C.-W. Luo *et al.*, npj Quantum Materials **2**, 32 (2017).

[19]     R. W. Hu, H. C. Lei, M. Abeykoon, E. S. Bozin, S. J. L. Billinge, J. B. Warren, T. Siegrist, and C. Petrovic, Phys Rev B **83** (2011).

[20]     D. Chareev, E. Osadchii, T. Kuzmicheva, J. Y. Lin, S. Kuzmichev, O. Volkova, and A. Vasiliev, Crystengcomm **15**, 1989 (2013).

[21]     S. P. Weathersby *et al.*, Rev. Sci. Instrum. **86** (2015).

[22]     L. Rettig *et al.*, Struct Dynam-Us **3** (2016).

[23]     D. Fobes *et al.*, Phys Rev Lett **112** (2014).

[24]     C. Ma *et al.*, Epl-Europhys Lett **84** (2008).

[25]     B. A. Frandsen *et al.*, Phys Rev Lett **119** (2017).

[26]     J. L. Niedziela, M. A. McGuire, and T. Egami, Phys Rev B **86** (2012).

[27]     X. J. Wu *et al.*, Appl Phys Lett **90** (2007).

[28]     A. V. Muratov, A. V. Sadakov, S. Y. Gavrilkin, A. R. Prishchepa, G. S. Epifanova, D. A. Chareev, and V. M. Pudalov, Physica B **536**, 785 (2018).

[29]     O. Zachar and I. Zaliznyak, Phys Rev Lett **91** (2003).

[30]     I. A. Zaliznyak *et al.*, Phys Rev B **85** (2012).

[31]     A. Patz, T. Li, S. Ran, R. M. Fernandes, J. Schmalian, S. L. Bud'ko, P. C. Canfield, I. E. Perakis, and J. Wang, Nature Communications **5**, 3229 (2014).

[32]     E. W. Carlson and K. A. Dahmen, Nature Communications **2** (2011).

[33]     L. M. Nie, G. Tarjus, and S. A. Kivelson, P Natl Acad Sci USA **111**, 7980 (2014).

[34]    T. M. McQueen *et al.*, Phys Rev B **79** (2009)..

[35]    E. J. Kirkland, *Advanced computing in electron microscopy* (Plenum Press, New York, 1998).

FIG.1. (Color online). Decay of the Bragg intensity during the first 5 ps for <200>, <020> (a) and <400>, <040> (b) reflections measured with UED at 27 K. Solid blue curves are exponential decay fits for the experimental data. (c) The peaks' dynamics during 1000 ps. Inset shows the FeSe diffraction pattern. The red line in the inset shows the intensity profile integrated within the indicated frame. (d) Changes of the intensity profile, shown in the inset (c), after the pump pulse with respect to the profile of the unpumped sample. Red(blue) line corresponds to the changes averaged over the time frame A(B) highlighted in (c).

FIG. 2. (Color online). (a) PDF at 84 K with the fit assuming an orthorhombic structural model. (b) PDF at 300 K with the fit assuming a tetragonal structural model. Blue circles show the experimental data, red line is the fit to the respective model, green line shows the misfit. The plots contain green (Fe-Se), blue (Fe-Fe) and red (Se-Se) tick marks below the residual, which indicate the different unique pair distances from refining the respective models.

FIG. 3. (Color online). (a) Electron diffraction pattern at 300 K (b) Electron diffraction from the same area as (a) at 88 K. (c) Typical HRTEM image. (d) FFTs taken from the respective areas as shown in (c). The peaks forbidden by the orthorhombic and tetragonal symmetry are highlighted by red circles.

FIG. 4. (Color online). (a) Dynamics of *<080>*, *<800>* peaks at different temperatures for the incident fluence of 1.65 mJ/cm$^2$. Dynamics at different excitation fluences for the full measurement time range (b) and during the first 150 ps (c) at 27 K. The gray dashed line in (c) is a guide to eye. Insets show schematics of unequal atomic bonds at the corresponding time intervals.

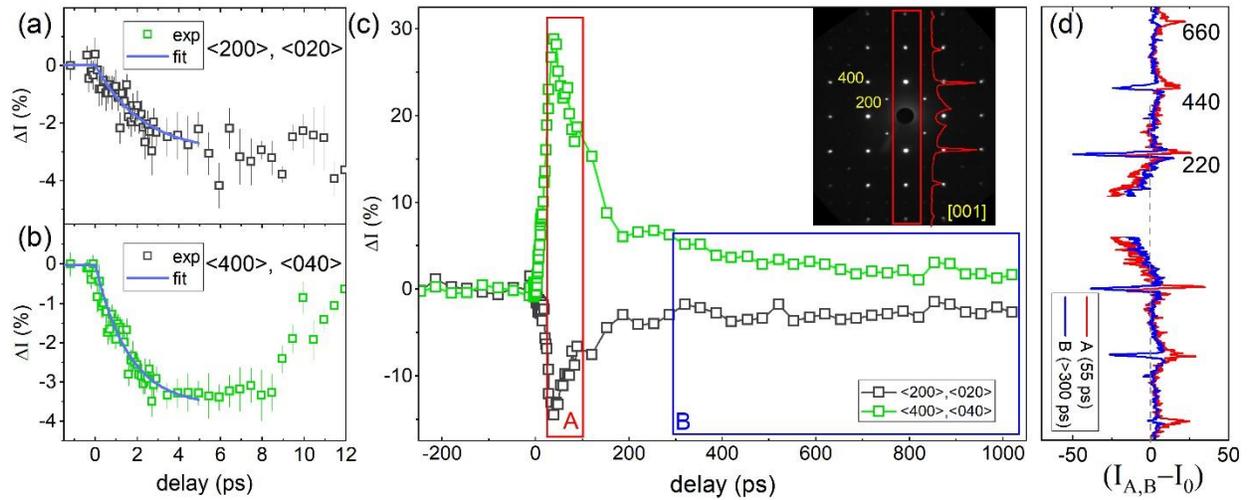

FIG.1. (Color online). Decay of the Bragg intensity during the first 5 ps for <200>, <020> (a) and <400>, <040> (b) reflections measured with UED at 27 K. Solid blue curves are exponential decay fits for the experimental data. (c) The peaks' dynamics during 1000 ps. Inset shows the FeSe diffraction pattern. The red line in the inset shows the intensity profile integrated within the indicated frame. (d) Changes of the intensity profile, shown in the inset (c), after the pump pulse with respect to the profile of the unpumped sample. Red(blue) line corresponds to changes averaged over the time frame A(B) highlighted in (c).

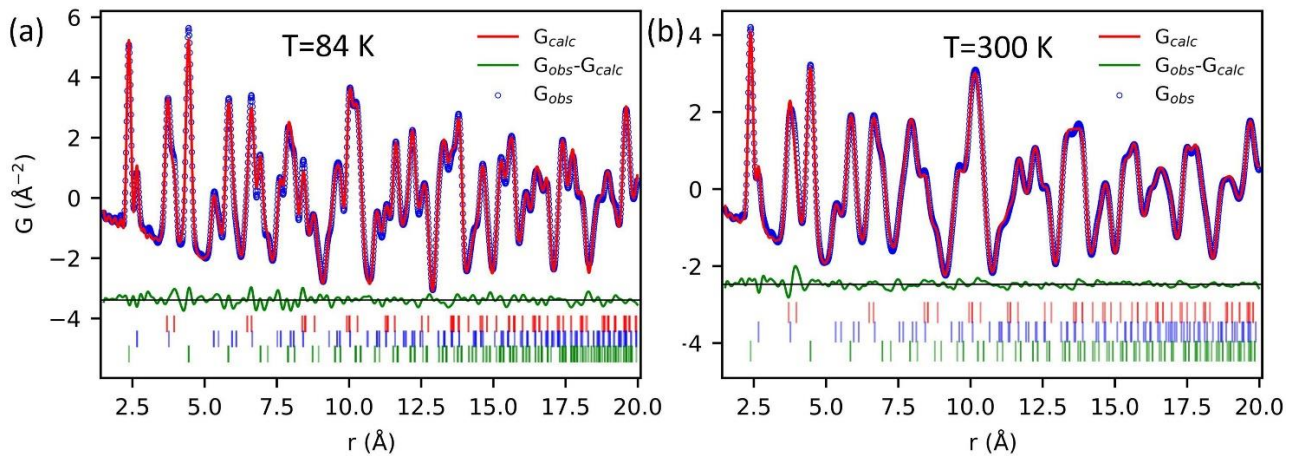

FIG. 2. (Color online). (a) PDF at 84 K with the fit assuming an orthorhombic structural model. (b) PDF at 300 K with the fit assuming a tetragonal structural model. Blue circles show the experimental data, red line is the fit to the respective model, green line shows the misfit. The plots contain green (Fe-Se), blue (Fe-Fe) and red (Se-Se) tick marks below the residual, which indicate the different unique pair distances from refining the respective models.

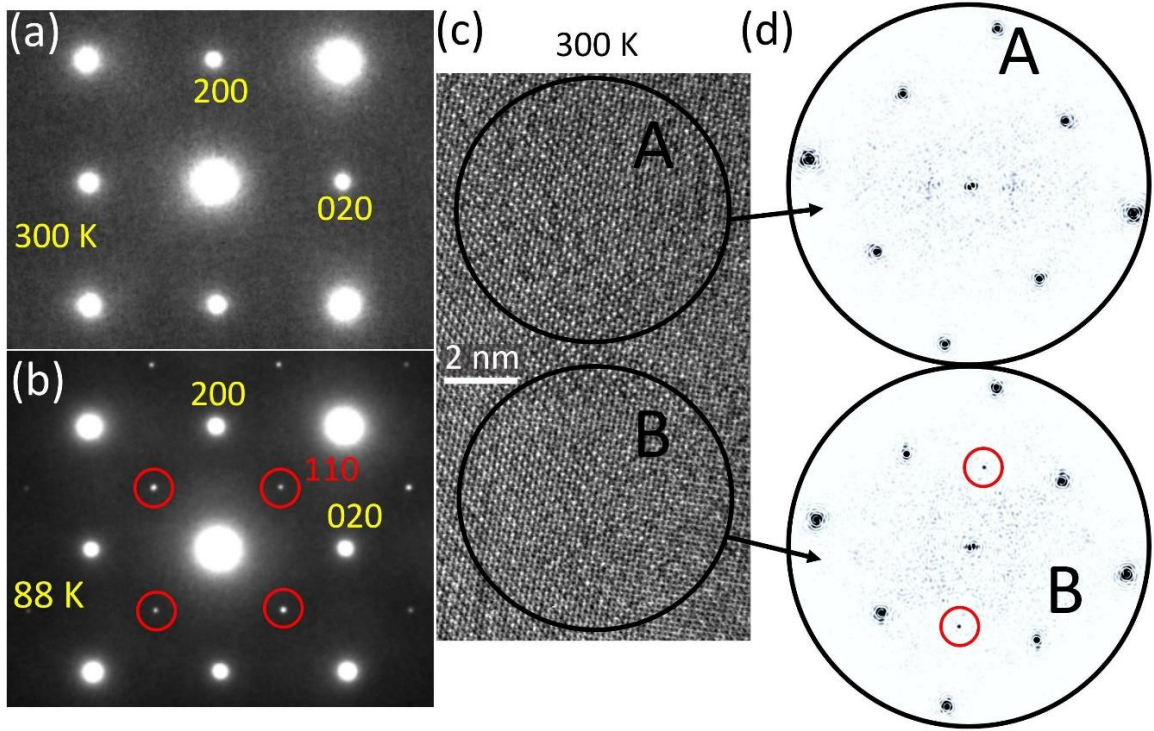

FIG. 3. (Color online). (a) Electron diffraction pattern at 300 K (b) Electron diffraction from the same area as (a) at 88 K. (c) Typical HRTEM image. (d) FFTs taken from the respective areas as shown in (c). The peaks forbidden by the orthorhombic and tetragonal symmetry are highlighted by red circles.

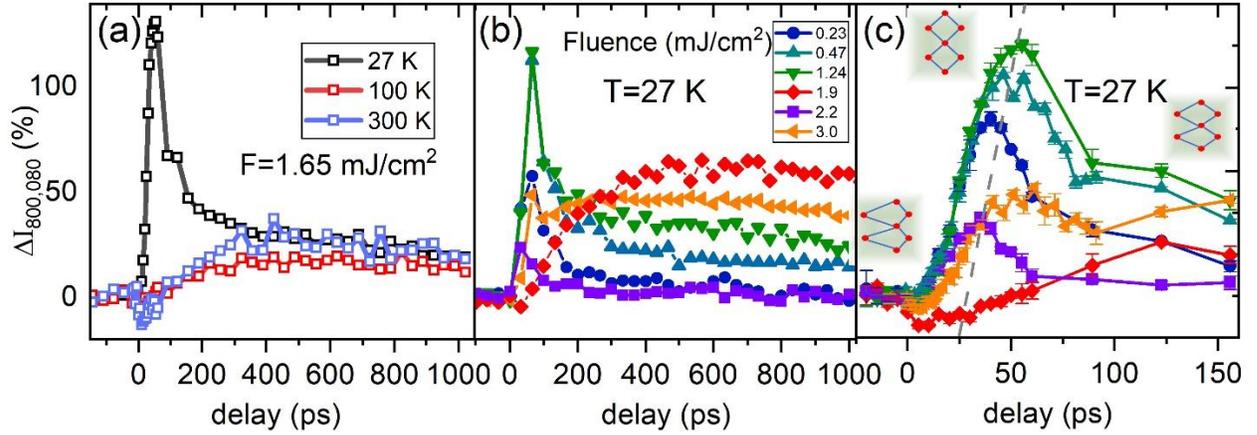

FIG. 4. (Color online). (a) Dynamics of *<080>*, *<800>* peaks at different temperatures for the incident fluence of 1.65 mJ/cm². Dynamics at different excitation fluences for the full measurement time range (b) and during the first 150 ps (c) at 27 K. The gray dashed line in (c) is a guide to eye. Insets show schematics of unequal atomic bonds at the corresponding time intervals.

# Photoinduced ultrafast dynamics of local nematicity and lattice distortions in FeSe crystals

T. Konstantinova[1,2], L. Wu[1], M. Abeykoon[1], R. J. Koch[1], A. F. Wang[1, §], R. K. Li[3], X. Shen[3], J. Li[1], J. Tao[1], I. A. Zaliznyak[1], C. Petrovic[1], S. J. L. Billinge[1,4], X. J. Wang[3], E. S. Bozin[1] and Y. Zhu[1,2]

[1] Brookhaven National Laboratory, Upton, NY, USA 11973
[2] Stony Brook University, Stony Brook, NY, USA 11794
[3] SLAC National Accelerator Laboratory, Menlo Park, CA, USA 94025
[4] Columbia University, New York, NY 10027, USA
[§] Present address: School of Physics, Chongqing University, Chongqing 400044, China

## 1. XPD measurements and Rietveld refinement

40-micron powder is prepared from FeSe single crystals. The powder is enclosed in evacuated glass capillaries. At BNL X-ray source with wavelength λ=0.1667 Å is used. Measurements at APS are done using λ=0.4128 Å source. Low temperature measurements are conducted using cryostream cooling.

Previous XPD analysis[34] determined the crystal symmetry as Cmma below $T_S$ = 91 K and P4/nmm above $T_S$. Rietveld refinement of the XPD data presented here agrees with the previous works. The powder diffraction data at 84 K can be fitted well using Cmma group with Fe atoms at 4$a$ (0.25, 0,0) sites and Se atoms at 4$g$ (0, 0.25, z). At 99 K the data can be fitted using P4/nmm group with Fe atoms at 2$a$ (0.75, 0.25, 0) and Se atoms at 2$c$ (0.25, 0.25, 0) sites. The data obtained at BNL at 84 K and the Rietveld refinement are shown in Fig. S1. The refinements at 84 K and 99 K are summarized in the Table S1.

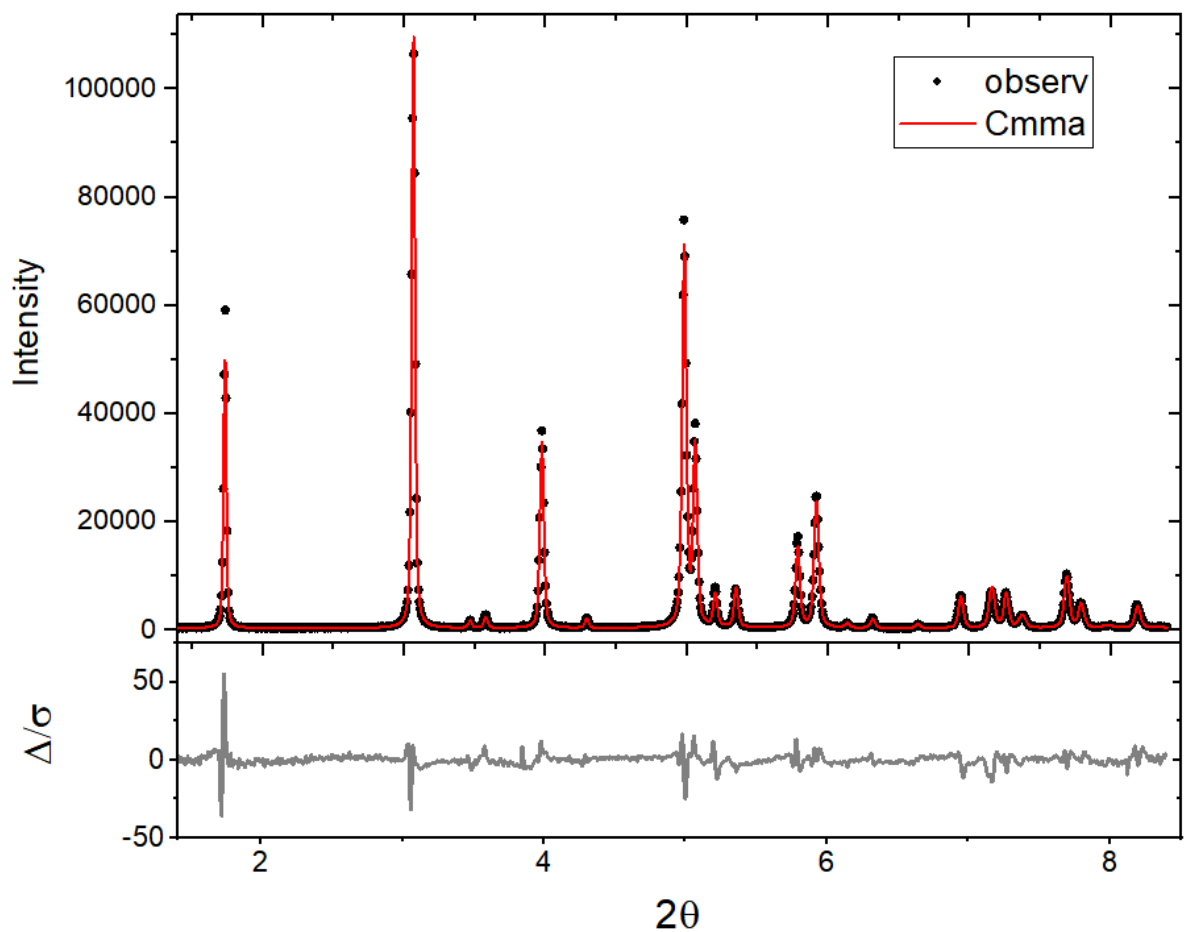

Fig. S1. XPD data at 84 K and Rietveld refinement using Cmma symmetry.

Table S1. Parameters of the Rietveld refinement of the XPD data. Measurement errors are indicated in brackets.

| T, K | Group symmetry | $a$ | $b$ | $c$ | $U_{Fe}$ | $U_{Se}$ |
|------|----------------|-----------|-----------|-----------|-----------|-----------|
| 84   | Cmma           | 5.3267(4) | 5.3120(4) | 5.4854(2) | 0.0069(7) | 0.0056(5) |
| 99   | P4/nmm         | 3.76176(8) |          | 5.4868(2) | 0.0053(7) | 0.0046(5) |

As discussed in the main text, the XPD data do not have peaks forbidden by the orthorhombic (or tetragonal) symmetry, such as <*110*> at any measured temperature. Fig. 2S shows the high-resolution XPD data obtained at APS for 300 K and 90 K with a closer look at the region where <*110*> peaks should be located based on the unit cell parameters. At the location there is no pronounced peak above the noise level. The splitting of the peaks at 90 K confirms the transition of the sample to the orthorhombic phase.

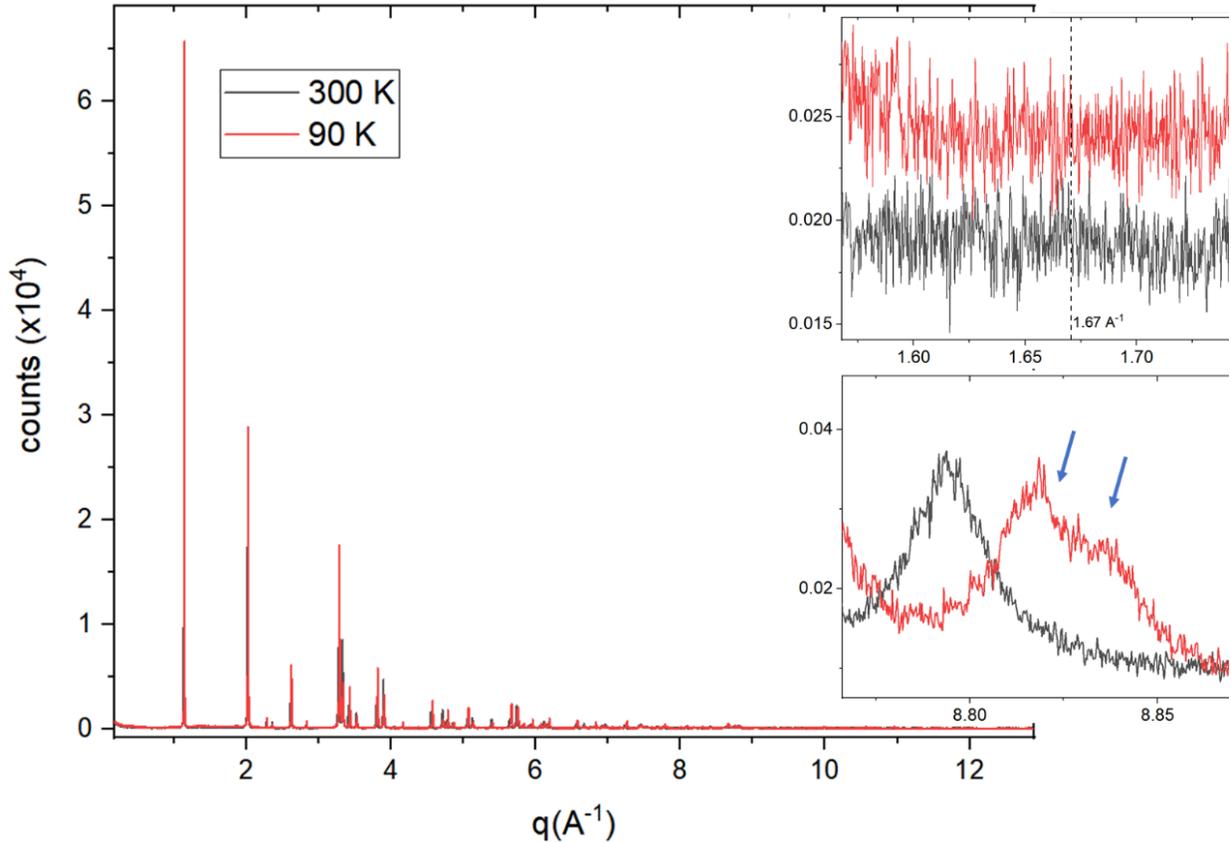

Fig. 2S. High-resolution XPD data at 90K and 300 K. Upper inset shows zoomed region near expected position of <*110*> peak. Position of reciprocal vector of <*110*> peak ($q_{110} = 1.67 Å^{-1}$) is marked with dashed line. Lower inset shows region at large diffraction angles where peak splitting associated with orthorhombic phase is well pronounced. The splitting is marked with arrows.

## 2. TEM/UED sample preparation

TEM / UED samples are prepared by exfoliation from a bulk single crystal. At first, a several microns thick part of the crystal, exfoliated with a sticky tape, is glued to a holder with wax. The exfoliation from this part continues until the remaining glued crystal becomes partially transparent for white light (tested with an optical microscope). The wax is then dissolved in acetone and the film is transferred to a commercially available nickel grid. The thickness of the samples is checked in TEM using electron energy loss spectroscopy and ranges from 10 to 150 nm. The typical thickness of the samples used in UED experiments is around 100 nm.

### 3. Separation of the peak profile into components.

For UED data it is possible to separate profiles of the intense Bragg peaks, such as *<040>*, *<220>*, *<620>*, etc., into two components: a Gaussian with a smaller width and a Lorentzian with a larger width, as shown in Fig. S3.

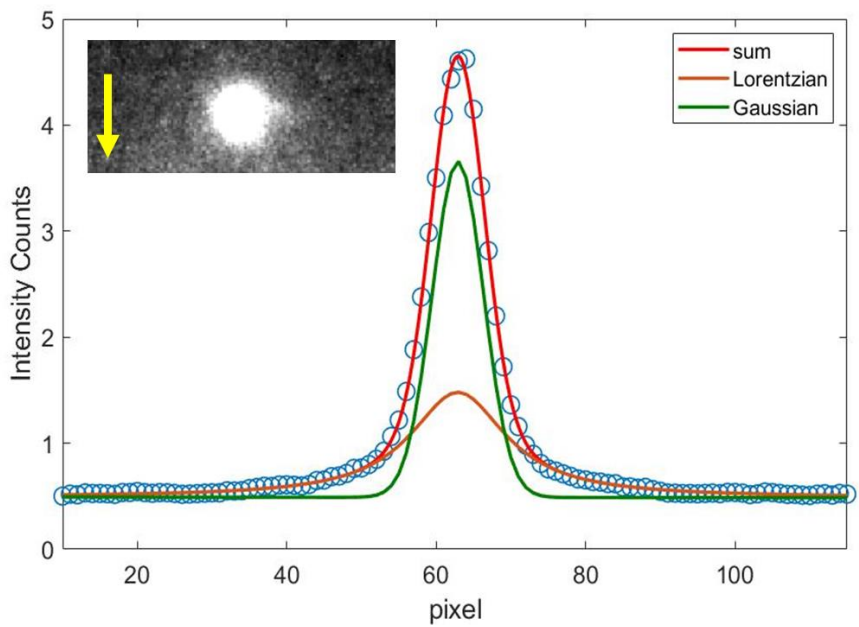

Fig. 3S. Profile of the *220* peak fitted with Gaussian (green) and Lorentzian (orange) components. Open circles are the experimental data. Inset shows the image of the window, within which the *220* peak profile is integrated along the axis indicated by arrow.

Our analysis shows (Fig. 4S) that the Gaussian and the Lorentzian parts have distinct behavior upon photoexcitation. For some peaks, e.g. *<400>* and *<040>*, both components follow the same trend, but the relative amplitude of change for the Lorentzian component is larger than for the Gaussian one. On the other hand, for *<220>* peaks, the Gaussian and Lorentzian component behave in the opposite ways. Whereas the amplitude of the Lorentzian component goes up after the pump pulse, the amplitude of the Gaussian component always decreases with time. The difference in behavior is explained by different nature of those components. The Gaussian part reflects the long-range order Bragg peaks, convoluted with the instrumentation profile. The change is this component is driven by the change of the lattice structure factors due to disorder (Debye-Waller effect) and by atomic displacements of atoms toward their high-symmetry positions. The Lorentzian part corresponds to scattering from a set of uncorrelated nanodomains. Due to the small signal-to-noise ratio for the weak ($h+k=4n+2$) peaks we are unable to separate their profiles into two components.

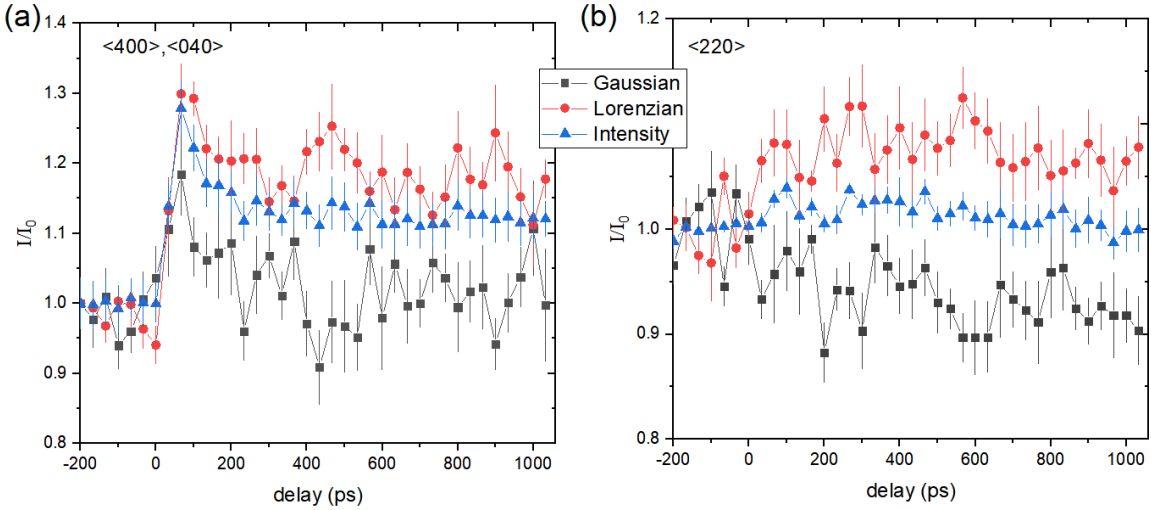

Fig. S4. Dynamics of Lorentzian and Gaussian components of *<400>*, *<040>* peaks (a) and *<220>* peaks (b) for sample at 27 K, excitation fluence 1.24 mJ/cm$^2$.

Note, that for the bright ($h+k=4n$) peaks the total intensity, which is the sum of both components, grows with respect to the value before the arrival of the pump pulse. Thus, the intensity does not simply redistribute between the components, but rather comes from other parts of the diffraction pattern, in particular, from the diffuse background centered at $q$=0. While the central beam is not recorded by the detector in our UED setup to avoid the detector oversaturation, we look at the dynamic of the intensity in the recorded area near $q$=0 and compare it to the dynamics of the Bragg peaks. Fig. S5. Shows that the dynamics are complimentary.

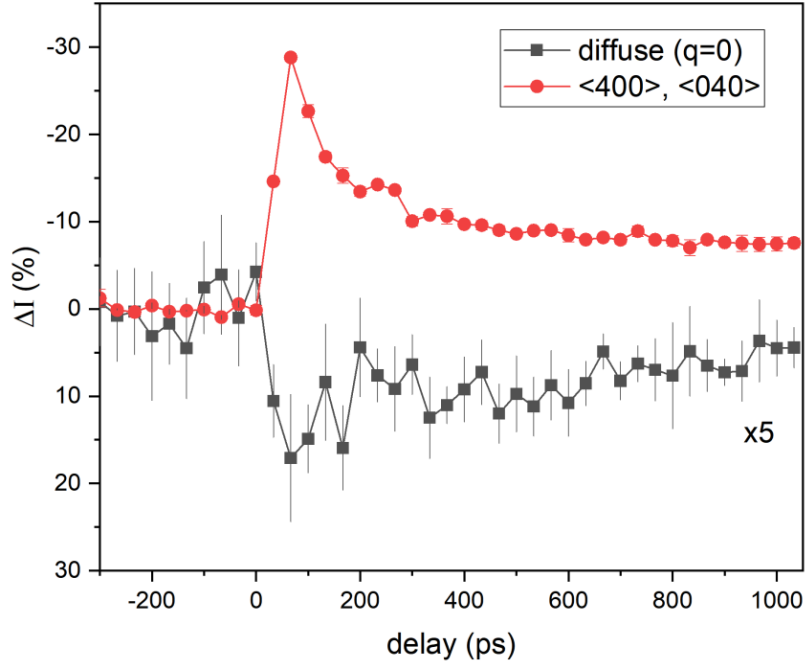

Fig. S5. Comparison of the dynamics the *<040>*, *<400>* Bragg peaks intensity and the diffuse intensity near $q$=0. Sample is at 27 K, excitation fluence is 1.24 mJ/cm$^2$.

### 4. UED experiments for FeSe$_{0.9}$S$_{0.1}$ sample.

UED measurements have also been performed for S-doped samples (FeSe$_{0.9}$S$_{0.1}$, T$_S$ = 70 K) under the excitation conditions similar to the ones described in the main text. The lattice response shown in Fig. S6 resembles the behavior of FeSe samples. Namely, at low fluences, after the initial 5 ps drop the signal begins to increase for $h+k = 4n$ peaks and continues to drop further for $h+k = 4n+2$ peaks for about 50 ps. Then, a slow recovery follows. As shown in Fig. S6(a), the amplitude of the fast rise of intensity for *<080>*, *<800>* peaks initially increases with the pump fluence and then begins to drop above 1.1 mJ/cm2. The intensity level at 1 ns delay continues to rise monotonically with fluence. When the sample temperature increases (Fig. S6(b)), the amplitude of the fast process at 50 ps goes down until it is no longer observed at temperatures above T$_S$.

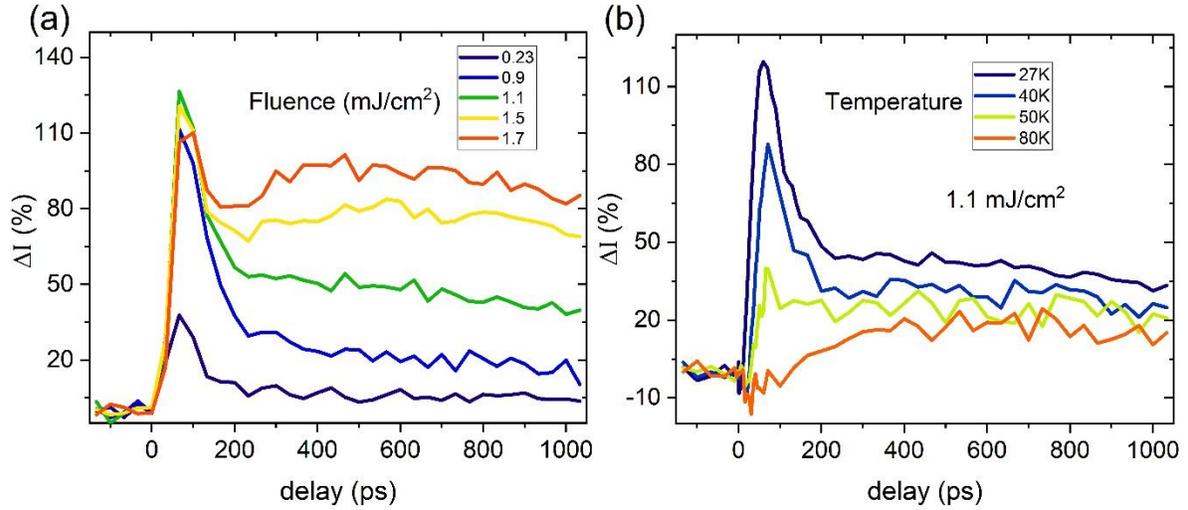

Fig. S6. Nonequilibrium lattice dynamics of $FeSe_{0.9}S_{0.1}$ sample. (a) Fluence dependence of *<080>*, *<800>* peaks dynamics. (b) Temperature dependence of *<080>*, *<800>* peaks dynamics.

Similarly to the pure FeSe, the dynamics in the doped $FeSe_{0.9}S_{0.1}$ sample are also characterized by intensity redistribution between the Bragg peaks and diffuse background. A convenient way to illustrate the intensity transfer is through the difference images. Figure S7 shows the changes in the UED pattern at +3.5, +55 and +1022 ps after the photoexcitation with the respect to the diffraction of the unpumped sample. The sample temperature is 27 K and the pump fluence is 1.1 mJ/cm2. As described earlier, at +3.5 ps delay the intensity of all Bragg peaks goes down from the initial value due to increased atomic vibrations. This intensity almost uniformly distributes in the background between peaks. At +55 ps delay the structural changes become apparent in the diffraction. The intensity of most $h+k = 4n$ peaks increases by taking up the intensity from the center of the diffraction pattern. At the same time, intensities of $h+k = 4n+2$ peaks are below their values before pulse. At +1022 ps the intensity of the peaks partially recovers from the values at +55 ps. The same is true for the diffuse intensity centered at $q$=0.

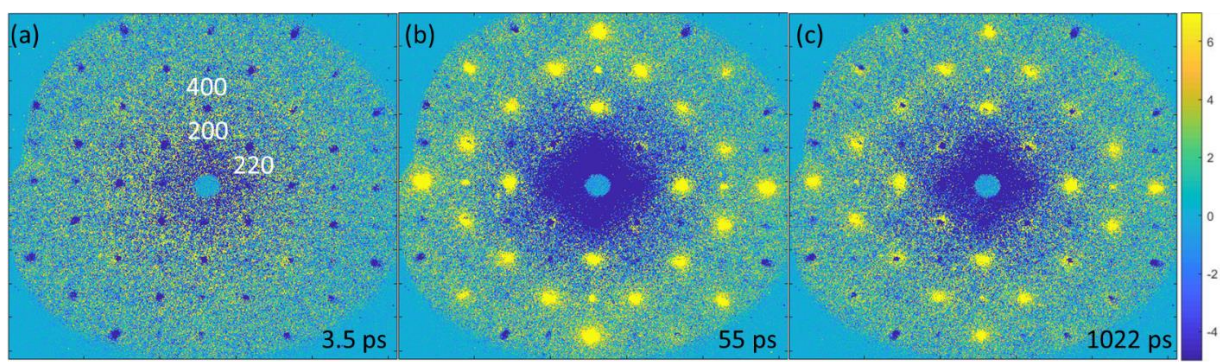

Fig. S7. Intensity transfer at different time delays. Difference diffraction patterns at +3.5 ps(a), +55 ps(b) and +1022 ps(c). Color range is the same for all figures and is encoded in the right panel.

## 5. Unequal population of twisted domains measured by TEM electron diffraction at 88 K

Analysis of the diffraction peak intensities along [110] and [1$\bar{1}$0] directions shows (Fig. S8) that while intensities of the <220> Bragg peaks are comparable along these directions, the intensity of the <110> peaks are different. This can be explained by unequal population of 90 degree domains with $C_2$ symmetry in the probed volume. Each domain produces peaks only along one of the directions, as observed in HRTEM images at 300 K. Probing multiple domains simultaneously results in appearance of all four peaks, where the peaks intensity depends on the population of each type of domain in the probed volume.

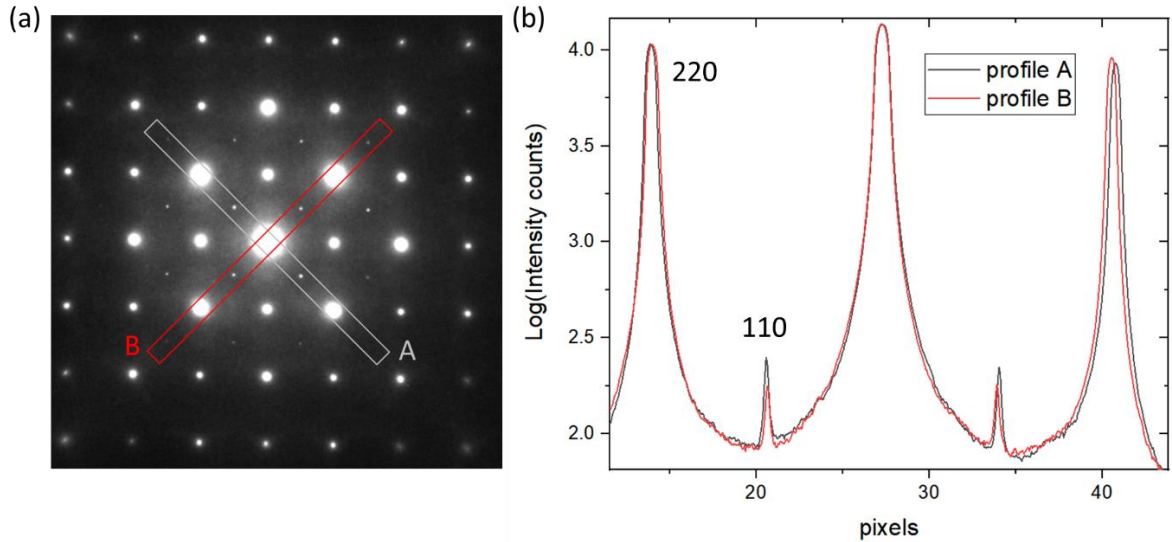

Fig. S8. Unequal peak intensities along two perpendicular directions. (a) Diffraction pattern of FeSe at 88 K. Grey and red boxes shows the windows within which the profiles were taken. (b) Comparison of intensities (on the log scale) integrated within widows A and B. A slight shift of the peaks on the right is due to the distortion of the microscope lens.

## 6. Scanning Transmission Electron Microscopy (STEM) data

STEM imaging is performed at different areas of the FeSe sample at 300 K. The Fourier Transform of the image shows the *<110>* peaks forbidden by both Cmma and P4/nmm symmetry in analogy to the HRTEM images. The peaks indicate the local lattice symmetry breaking. An example of the STEM image and its Fourier Transform is shown in Fig. S9. As can be seen, intensities of the 110, $\bar{1}\bar{1}0$ peaks are higher than for $1\bar{1}0$ and $\bar{1}10$, pointing to unequal population of twisted domains in the probed area.

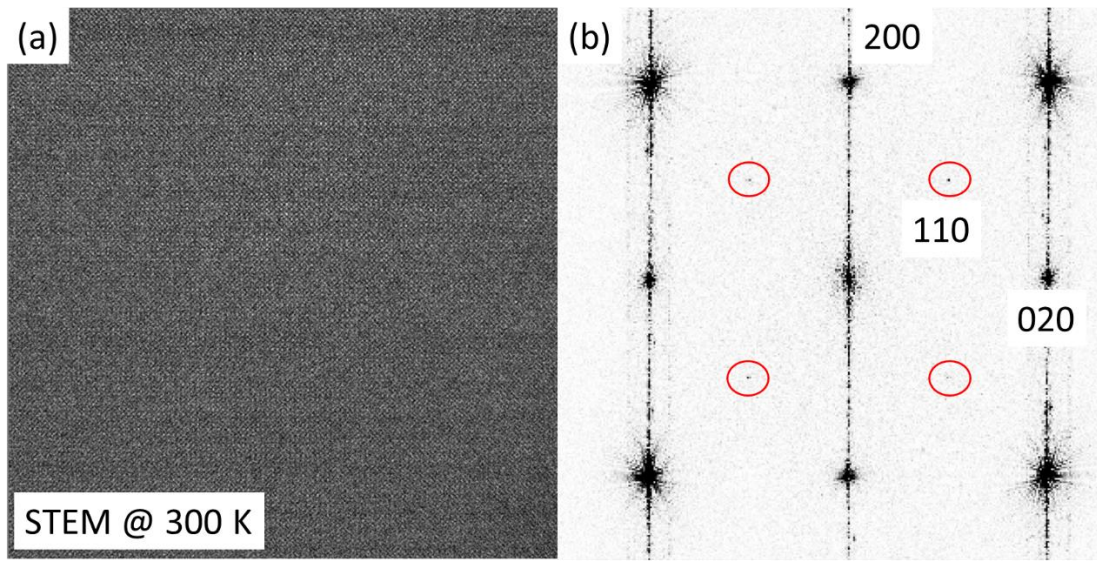

Fig. S9. Cmma symmetry breaking detected by STEM. (a) STEM image of FeSe sample at 300 K. (b) The Fourier Transformation of (a). Forbidden *<110>* peaks are highlighted with red circles. Vertical lines are the artifacts caused by sample drift during scanning.

## 7.  Analysis of HRTEM images

As discussed in the main text, at 300 K HRTEM images of FeSe single crystals reveal sample areas with broken lattice symmetry. Additional examples of the Fourier analysis for HRTEM images are shown in Fig. S10.

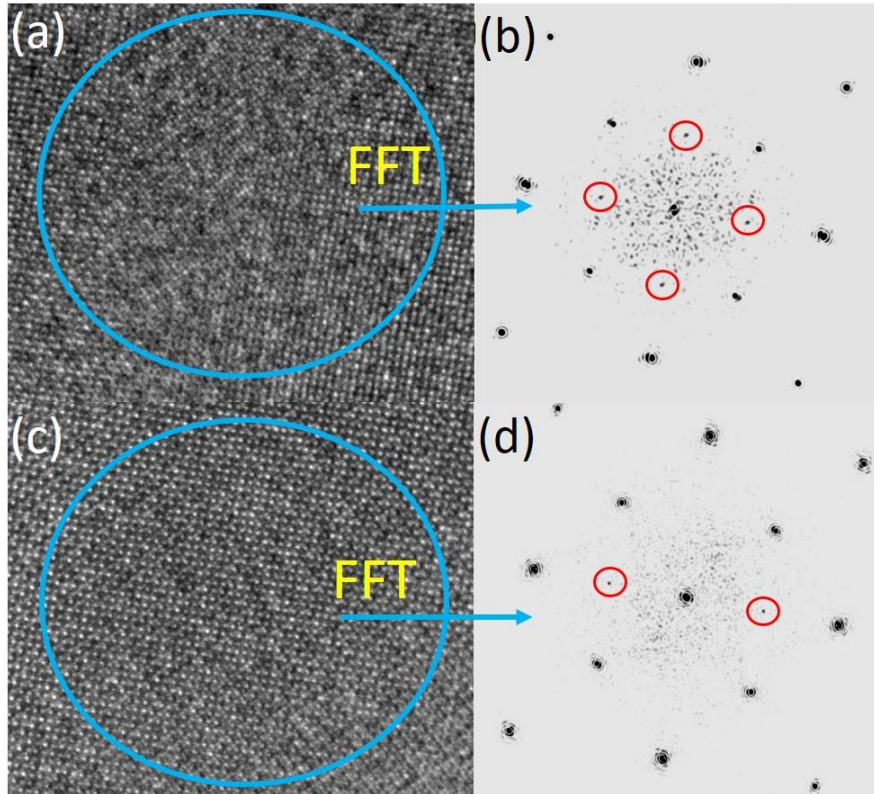

Fig. S10.  Analysis of HRTEM images of FeSe at 300 K. (a), (c) HRTEM images taken from different parts of the sample. (b), (d) The respective FFT of (a) and (c). Peaks, corresponding to the lower symmetry are highlighted with red circles.

In the UED experiments the extra peaks are observed for only part of the samples and their appearance can be explained by multiple scattering effects involving the first order Laue zone peaks. In such cases, the intensities of the extra peaks become stronger at large scattering angles and do not change with temperature.

The absence of the peaks, observed in the Fourier transformed HRTEM images at 300 K, in electron diffraction data can be explained by different sensitivities of the techniques to a weak atomic potential. The intensity of the Fourier transform of a HRTEM image is linearly proportional to the atomic potential, whereas the intensities of the Bragg peaks in diffraction are proportional to the squared potential[35]. Since electron diffraction does not detect the lower crystal symmetry at high temperatures we conclude that the coherence length of the distortions is very small. This agrees with the PDF data, where misfit to a tetragonal model is pronounced at small $r$ only. As the temperature of the sample is decreased below $T_S$, the correlation length of the distortion grows, and the distortions become detectable with electron diffraction in TEM. There are several reasons why the extra peaks, reflecting low symmetry, are not seen in the x-ray and UED data. Firstly, the intensities of the peaks are very weak and may not exceed the noise level in the XPD experiments. Besides, the signal comes from the structures with a limited coherence length even at low temperatures and requires a high degree of coherence of the diffracting beam, that may not be achieved in these experiments. Equivalently, we can say that the x-ray and UED probes view the atomic displacements as a random disorder rather than a systematic modulation

## 8. Dynamics of *<200>*, *<020>*, *<400>* and *<040>* peaks at different sample temperatures and pump fluences.

The behavior of *<200>*, *<020>*, *<400>* and *<040>* peaks is consistent with the dynamics of *<800>*, *<080>* peaks at the same excitation conditions. The difference in behavior of *<200>* and *<400>* peaks can be understood from their structure factor changes. Let $x_{Fe} = 0.25 + \delta x_{Fe}$ and $x_{Se} = 0 + \delta x_{Se}$ be the distorted coordinates of Fe and Se atoms along the a-axis respectively, where $\delta x_{Fe} \ll 1$ and $\delta x_{Se} \ll 1$ are the amplitudes of the distortions. Then the structure factors for *<200>* and *<400>* would be:

$$SF_{200} = 4f_{Se}B_{Se}\cos(4\pi \, \delta x_{Se}) - 4f_{Fe}B_{Fe}\cos(4\pi \, \delta x_{Fe}) \qquad (S1)$$

$$SF_{400} = 4f_{Se}B_{Se}\cos(8\pi \, \delta x_{Se}) + 4f_{Fe}B_{Fe}\cos(8\pi \, \delta x_{Fe}) \qquad (S2)$$

As in the main text, $f_{Se}$ ( $f_{Fe}$ ) and $B_{Se}$ ($B_{Fe}$) are atomic form factor and Debye-Waller factor for Se (Fe) atoms respectively. When the distortions are released, i.e. $|\delta x_{Se}|$ and $|\delta x_{Fe}|$ become smaller the amplitude of *<400>* peak increase, while the amplitude of *<200>* peaks can go up or down, depending on the relative change of $\delta x_{Se}$ and $\delta x_{Fe}$. Considering that the scattering power of Fe atoms is weaker than of Se atoms, the Fe displacements $\delta x_{Fe}$ change more upon photoexcitation than the displacements of Se atoms $\delta x_{Se}$. The same logic applies for $<020>$, $<040>$ peaks.

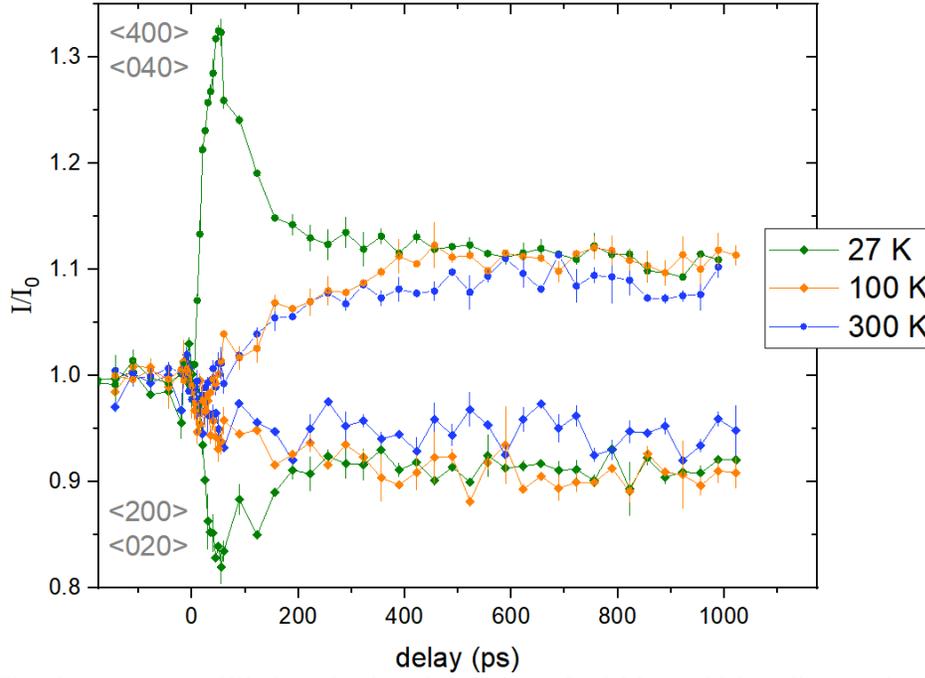

Fig. S11. Nonequilibrium lattice dynamics of $<200>$, $<020>$ (diamonds), $<400>$ and $<040>$ (circles) peaks for pure FeSe sample at different sample temperatures.

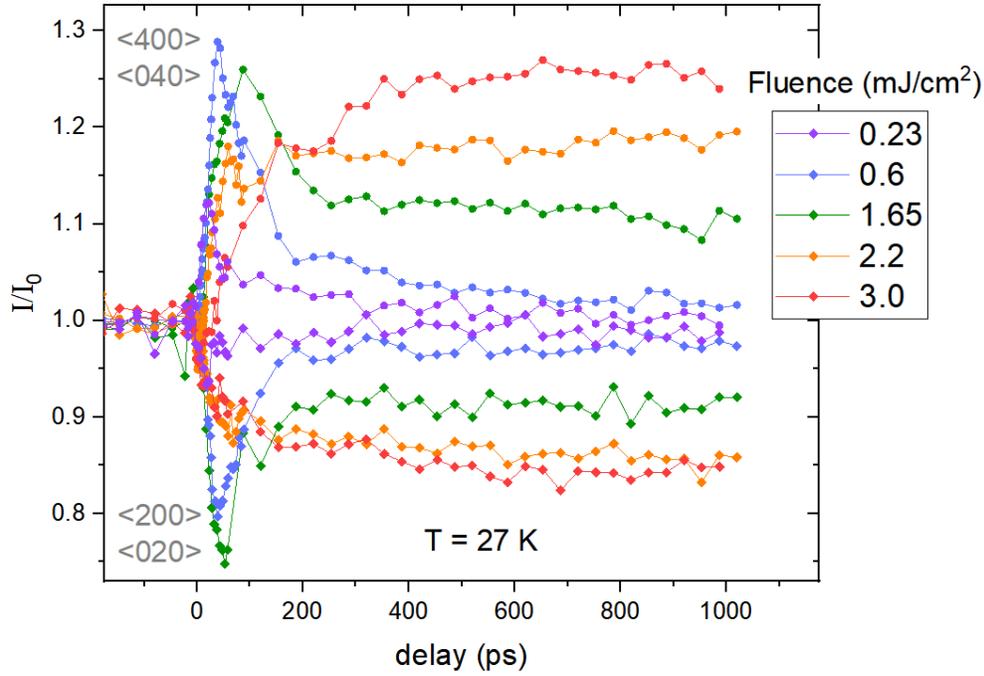

Fig. S12. Nonequilibrium lattice dynamics of *<200>*, *<020>* (diamonds), *<400>* and *<040>* (circles) peaks for pure FeSe sample at different pump fluences.

### 9. Comparing the experimental conditions among present and previous studies.

Both fast release of the distortions and orthorhombic-to-tetragonal lattice phase transition[22] in FBSCs proceed with a rate 100-200 times slower than the melting rate of primary electronic nematic order[18,31]. The decay of Bragg peaks' intensities after 50 ps at low temperatures is associated with partial reestablishment of the lower lattice symmetry but also with fracturing of the long-range-ordered regions into smaller nano-size domains. Again, such process is considerably slower than the recovery[18,31] of the electronic nematic order. Note that despite the observed slowing of the structural transition process with fluence, the minimum incident pump fluence used in our UED experiment(0.23 mJ/cm$^2$) is comparable to the pump fluence used[31] to measure the electronic response (0.17 mJ/cm$^2$). Accordingly, the contrast between the electronic and the lattice responses is unlikely caused by the differences in the excitation regimes.